\documentstyle[11pt,newpasp,psfig,twoside]{article}
\index{instructions}
\index{guidelines}

\def\edcomment#1{\iffalse\marginpar{\raggedright\sl#1\/}\else\relax\fi}
\reversemarginpar

\begin{document}
\title{The Galactic population and birth rate of radio pulsars}
\author{D.R.~Lorimer}
\affil{University of Manchester, Jodrell Bank Observatory, UK}

\begin{abstract}
We review current understanding of the underlying, as opposed to the
observed, pulsar population.  The observed sample is heavily biased by
selection effects, so that surveys see less than 10\% of all
potentially observable pulsars.  We compare various techniques used to
correct the sample for these biases. By far the most significant
recent development has been the discovery of over 700 pulsars in the
Parkes Multibeam (PM) survey. This new sample is far less affected by
selection effects and we use it to make a preliminary analysis of the
Galactic pulsar distribution, finding further evidence for a deficit of 
pulsars in the inner Galaxy.
\end{abstract}

\section{Selection effects in pulsar surveys}

\begin{figure}[hbt]
\psfig{file=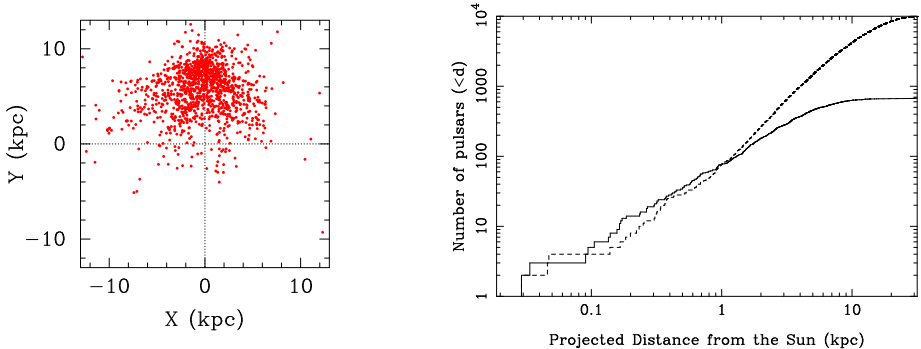,width=\textwidth}
\caption{\label{fig:selfx}
Left: The currently known pulsar population projected onto the
Galactic plane. The Galactic centre is at the origin and the Sun is at
(0.0,8.5) kpc. Right: Cummulative distribution of pulsars as a
function of distance from the Sun projected onto the plane. The solid
line shows the observed sample (circa 1997) while the dashed line shows the
expected distribution of a population free from selection effects.}
\end{figure}

Fig.~\ref{fig:selfx} shows the current sample of 1300 pulsars in the
ATNF on-line catalogue (Hobbs et al.~these proceedings) projected onto
the Galactic plane.
Rather than being distributed about the Galactic centre, the majority
of pulsars are local objects.  Far from being representative of the
true population, this sample is heavily biased by a number of
selection effects which we now outline below.

\noindent
{\bf The inverse square law.} Like all astronomical sources, observed
pulsars of a given luminosity $L$ are strongly selected by their
apparent flux density, $S$.  For pulsars, which beam to a certain
fraction $f$ of $4\pi$ sr\footnote{In the absence of prior knowledge
about beaming, geometrical factors are usually ignored and the
resulting ``pseudoluminosity'' is quoted at some standard observing
frequency; e.g., at 1400 MHz, given a measured flux density and
distance, the pseudoluminosity $L_{1400} = S_{1400} d^2$.}  , $S =
L/(f4\pi d^2)$, where $d$ is the distance to the pulsar.  Since all
pulsar surveys have some limiting flux density only those objects
bright or close enough will be detectable.

\smallskip
\noindent {\bf The radio sky background.} One limit to pulsar search
sensitivities is the thermal noise in the receiver, i.e.~the ``system
temperature'', $T_{\rm sys}$. While every effort is made to minimize
$T_{\rm sys}$ at the telescope, synchroton radiating electrons in the
Galactic magnetic field contribute significantly with a ``sky
background'' component, $T_{\rm sky}$. At observing frequencies $\nu
\sim 0.4$ GHz, $T_{\rm sky}$ dominates $T_{\rm sys}$ along the
Galactic plane.  Fortunately, $T_{\rm sky} \propto \nu^{-2.8}$ so this
is effect is significantly reduced at higher frequencies; e.g.~for the
PM system, $\nu = 1.4$ GHz.

\begin{figure}[hbt]
\psfig{file=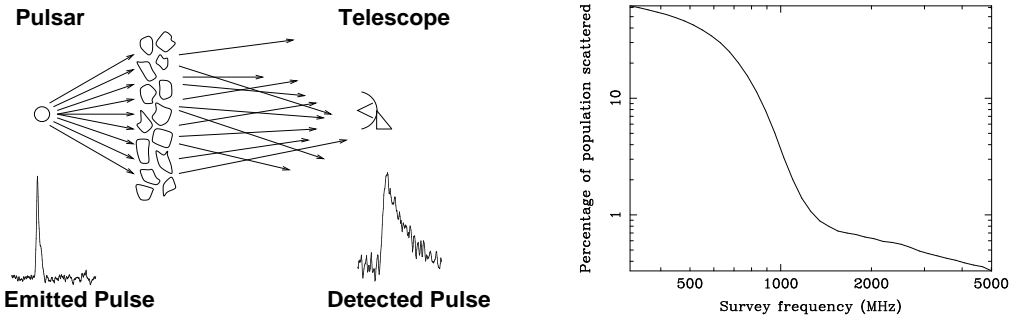,width=\textwidth}
\caption{\label{fig:scatt}
Left: The effect of pulse scattering by irregularities in the ISM.
Right: A simulation showing the estimated
fraction of pulsars rendered invisible
by scattering as a function of observing frequency.}
\end{figure}

\smallskip
\noindent
{\bf Propogation effects in the intersellar medium (ISM).} The ISM is
a mixed blessing for pulsar astronomers. On one hand, the dispersion
of pulses caused by the ionized component of the ISM permits an
estimate of $d$ through the dispersion measure. Conversely, dispersion
and scatter-broadening of the pulses conspire against detection of
short period and/or distant objects. The effects of scattering are
shown in Fig.~\ref{fig:scatt}.  Fortunately, like $T_{\rm sky}$, the
scatter-broadening time $\tau_{\rm scatt}$ has a strong frequency
dependence, scaling roughly as $\nu^{-4}$.
As shown in Fig.~\ref{fig:scatt}, there is a transition frequency of
about 1 GHz, below which scattering can hide a large fraction of the
population. Another factor is scintillation, the modulation of
apparent flux densities by refractive or diffractive ``screens'' of
material along the line of sight (Rickett 1970).  This is particularly
important for nearby pulsars, where the apparent flux densities can
vary significantly. For example, two northern sky surveys carried out
20 years apart with comparable sensitivity (Damashek, Taylor \& Hulse
1978; Sayer, Nice \& Taylor 1997) detected a number of pulsars above
and below the nominal search thresholds of one experiment but not the
other.  Ideally, surveying the sky multiple times minimizes the
effects of scintillation against pulsars nominally above the
threshold, and maximizes the detections of faint pulsars through
favourable scintillation.

\smallskip
\noindent {\bf Finite size of the emission beam.} As mentioned above,
the fact that pulsars do not beam to $4\pi$ sr means that we see only
a fraction $f$ of the total active population. For a circular pencil
beam, Gunn \& Ostriker (1970) estimated $f \sim 1/6$.  A consensus on
the precise shape and evolution of the emission beam, however, has yet
to be reached. Narayan \& Vivekanand (1983) argued that the beam shape
is elongated in the meridonial direction. Lyne \& Manchester (1988),
on the other hand, favour a circular beam. Using the same database,
Biggs (1990) presented evidence in favour of meridonal compression!
All of these studies do agree that the beam size is period dependent,
with shorter period pulsars having larger beaming fractions. For
example, Tauris \& Manchester (1998) found that
$f=0.09\left[\log(P/{\rm sec})-1\right]^2+0.03$, where $P$ is the
period.  Undoubtedly, a complete model for $f$ needs to account for
other factors, such as evolution of the inclination angle between the
spin and magnetic axes. Given the uncertainties, most authors quote
results with and without a beaming correction.

\smallskip
\noindent {\bf Pulse nulling.} The abrupt cessation of the pulsed
emission for many pulse periods, was first identified by Backer
(1970). Ritchings (1976) presented evidence that the incidence of
nulling became more frequent in older long-period pulsars, suggesting
that it signified the onset of the final stages of the neutron star's
life as an active radio pulsar. Since most pulsar surveys have short
($<$ few min) integration times, there is an obvious selection effect
against nulling objects. Means of combatting this effect are to look
for individual pulses in search data (see e.g.~Nice 1999), survey the
sky many times, or use longer integrations. The PM survey, which
employs 35-min pointings, is proving particularly effective at
detecting nulling pulsars and should soon be able to better quantify 
this population and provide a more satisfactory understanding of 
nulling pulsars.

\smallskip
\noindent {\bf Orbital motion.} In standard pulsar searches, where
time series are Fourier transformed, the signal from a binary pulsar
can be Doppler shifted over several bins in the Fourier domain. In
extreme cases, where the survey integration time is a significant
fraction of the orbital period, this results in a loss of sensitivity.
For example the loss of signal-to-noise of the original binary pulsar
B1913+16 during a 35-min observation of the PM survey can be as much
as 90\%. Correcting for this effect using
``acceleration searches'', is now becoming more routine, thanks to the
ever-increasing availability of high-speed computer resources. Since
it is quite possible that young pulsars in tight binary orbits exist,
the very deep ($\sim$ 2--10 hr) searches for young pulsars described
by Camilo in this volume are now being re-analysed with full
acceleration searches.

\smallskip
To get an idea of how biased the sample is due to the above effects,
Fig.~\ref{fig:selfx} shows the cummulative distribution of pulsars as
a function of distance from the Sun projected onto the Galactic
plane. Also shown is the expected distribution for a simulated
population in which there are no selection effects.  As can be seen,
the two samples are closely matched only out to a kpc or so before the
selection effects become significant. From these curves, we deduce
that {\it less than 10\% of the potentially observable population in
the Galaxy are currently detectable.} Rigorous conclusions about
the true pulsar population can only be made after properly accounting
for these selection effects.

\section{Landmark papers in pulsar statistics}

Shortly after the discovery of pulsars, a number of authors began to
consider their implications for the Galactic population of neutron
stars. A widely cited paper from that era is the work of Gunn \&
Ostriker (1970; hereafter GO).  To tackle the problem analytically, GO
made two simplifying assumptions based on the sample of 52 pulsars
known at that time: (a) the relationship $L \propto B^2$, where $L$ is
the radio luminosity, and $B$ is the dipole magnetic field strength;
(b) the evolution of $B$ being an exponential decay with a time
constant $t_d$. With these in hand, and using the dipolar spin-down
expression $B^2 \propto P \dot{P}$, GO derived expected
distributions for the observed population. In particular, they showed
that the observed number of pulsars $N_{\rm obs} = \pi D^2 F \Sigma
t_d \exp(2 \sigma_P^2)/2$, where $D$ is the mean distance of the
observed sample, $F$ is a parameter relating to the completeness of
the surveys, $\Sigma$ is the local birth rate of potentially
observable pulsars and $\sigma_P$ is derived from their model fit to
the observed period distribution. Assuming $f=1/6$, and extrapolating
the local birth rate over the Galaxy, they arrived at a Galactic birth
rate ${\cal R} = 1/(30 \, {\rm yr})$, in good agreement with the best
estimates of the rate of type-II supernovae at that time (e.g.~Blaauw 1961).

Although mathematically appealing, GO's analytical approach required a
number of simplifying assumptions about the pulsar population and the
Galaxy itself. One of these assumptions, the spontaneous decay of the
magnetic field, continues to be extremely controversial. Since GO's
original study, many papers have been written presenting arguments for
and against field decay (see e.g.~Lyne, Manchester \& Taylor 1985;
Bailes 1989; Narayan \& Ostriker 1990; Bhattacharya et al.~1992; see
also the contribution by van Leeuwen et al. in this volume). A less
model-dependent approach to the problem, first developed by Large
(1971), can be summarized by the following expression:
\begin{displaymath}
dn(P,z,R,L) = V(P,z,R,L) \rho(P,z,R,L) dP\,dz\,dR\,dL.
\end{displaymath}
Here $N$ is the observed population of pulsars as a function of
period, $P$, distance from the Galactic plane, $z$, Galactocentric
radius, $R$ and luminosity, $L$. The quantity $V$ represents the
volume of the Galaxy effectively searched and $\rho$ is the underlying
(true) distribution of the population. Since we know $n$ and can
estimate $V$ on the basis of (hopefully) well-understood survey
sensitivities, we can invert the above expression to solve for
$\rho$. The only simplification required to do this is to
assume\footnote{Following a question about this assumption after my
talk, I examined the current pulsar sample. Apart from a weak coupling
between $P$ and $z$, there are no significant relationships between
any of these parameters. The assumption of independent distributions
seems to be well founded.}  that $P$, $z$, $R$ and $L$ are independent
quantities. The problem then reduces to four equations which can be
solved for the underlying distributions of interest: $\rho_P(P)$,
$\rho_z(z)$, $\rho_R(R)$ and $\rho_L(L)$.

Large's method was somewhat ahead of its time: back in 1971, pulsar
surveys of the Galaxy were still in their early stages so that $V$ was
not well determined.  By the late 1970s, however, a number of
large-scale searches had been carried out and Taylor \& Manchester
(1977; hereafter TM) applied the above technique to the sample of
$\sim 150$ pulsars then known (see also Davies, Lyne \& Seiradakis
1977).  Integrating the derived distribution functions over the
Galaxy, and assuming $f=0.2$, TM estimated the total number of active
pulsars in the Galaxy to range between 60,000 and 850,000. Here the
principle uncertainty is the assumed distance model. TM considered a
uniform electron content with a mean electron density in the range
$0.02 < n_e < 0.03$ cm$^{-3}$.  To calculate the Galactic birth rate,
${\cal R}$, TM required an estimate of the mean pulsar lifetime, $T$,
which they obtained from an analysis of the $z$ distribution as a
function of characteristic age $\tau=P/(2\dot{P})$. In the $z-\tau$
diagram (see their Fig.~7), TM argued that the discrepancy between the
expected and observed $z$ for characteristic ages larger than a few
Myr set a limit to $T=4$ Myr.  This leads to an implied birthrate
${\cal R} = 1/(6 \,{\rm yr})$!  Davies et al.~(1977) reached similar
conclusions.

The rather high birth rates from these and other analyses prompted
Phinney \& Blandford (1981) and in particular Vivekanand \& Narayan
(1981; hereafter VN) to consider a less model-dependent approach to
estimating ${\cal R}$. The method involves binning the observed sample
as a function of spin period, $P$, and in each bin computing the flow
or ``current'' of pulsars, $J$, through the bin:
\begin{displaymath}
J(P) = \frac{1}{\Delta P} \sum_{i=1}^{n_{\rm bin}} \frac{\dot{P}_i \xi_i}{f_i}.
\end{displaymath}
Here, $n_{\rm bin}$ is the number of pulsars in a period bin of width
$\Delta P$, $\xi_i$ and $f_i$ are the ``scale factor'' and beaming
fraction of the $i^{\rm th}$ pulsar respectively. As discussed
earlier, $f_i$ is based on some beaming model. For a given pulsar, its
scale factor $\xi_i$ represents the number of pulsars with similar
parameters in the Galaxy. In principle, this is very similar to the
$V/V_{\rm max}$ first used to correct observationally biased samples
of quasars (Schmidt 1968). In practice $\xi$ is computed using a Monte
Carlo simulation of $N$ pulsars with identical periods and
luminosities. Using accurate models for the various pulsar surveys, it
is relatively straightforward to calculate the number of pulsars $n$
that are detectable from that population. As a result,
$\xi=N/n$. Detailed simulations to test this approach show that the
scale factors give reliable results about the population of pulsars
with luminosities above $L_{\rm min}$, the minimum luminosity in the
sample (see Lorimer et al.~1993).  We note in passing that it is
possible to form the luminosity function for $L>L_{\rm min}$ by
weighting the observed luminosities by the appropriate scale factors.
Integrating this distribution then yields $N(>L_{\rm min})$.

The beauty of the pulsar-current analysis is that it makes just two
fundamental assumptions about pulsars: (i) they are a steady-state
population; (ii) they are spinning down steadily from short to long
periods. The first of these is justified since the ages of pulsars,
while not well known ($10^{7-8}$ yr), are certainly less than the age
of the Galaxy, $10^{10}$ yr.  The second is, of course, well in accord
with timing observations.  The birth-rate can be computed from this
analysis by simply plotting $J$ as a function of $P$. In the standard
model where pulsars are born spinning rapidly, there should be a peak
in the current at short periods followed by a decline in the current
as pulsars end their life with longer periods. The birth rate is then
just the height of this peak.  VN derived ${\cal R} = 1/(16-27 \,{\rm
yr})$, in much better agreement with the supernova rate.

A by-product of VN's analysis was their conclusion that a significant
fraction of pulsars are ``injected'' into the population with
relatively long initial spin periods ($P_0 \sim 0.5$ s), rather than
the standard picture of birth with $P_0 \sim 20$ ms. Rather like field
decay, arguments for and against injection have been presented ever
since. Following criticism by Lyne, Manchester \& Taylor (1985;
hereafter LMT) that VN's analysis had not properly taken account of
selection effects such as scattering, Narayan (1987) performed a more
detailed analysis which provided further support for
injection. Lorimer et al.~(1993) used Monte Carlo simulations to
investigate the validity of the pulsar current approach and found that
it was prone to systematic errors induced by faint, nearby
pulsars. Excluding these objects from the analysis significantly
reduces the case for injection.  Perhaps the final word on this issue
will be a detailed pulsar current analysis of the PM survey sample
(Vranesevic et al.~this volume).

\begin{figure}[hbt]
\psfig{file=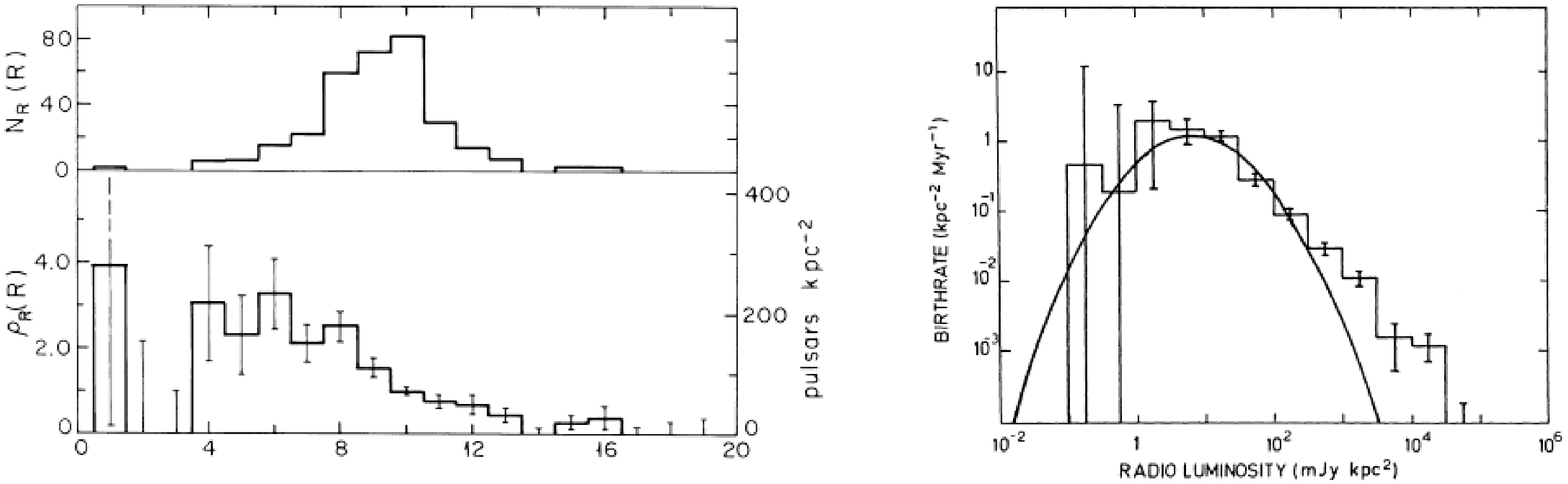,width=\textwidth}
\caption{\label{fig:lmt} Left: The observed radial distribution (top
panel) and corrected radial density function (lower panel) from the
numerical analysis of LMT. Right: The birthrate as a function of
luminosity assuming the Gunn--Ostriker luminosity model with dipolar
spindown and exponential magnetic field decay.  The solid line shows
the expected distribution of initial luminosities assuming a Gaussian
spread in magnetic field.}
\end{figure}

A major effort to quantify the Galactic distribution was the analysis
of LMT who combined the theoretical framework of GO with the numerical
approach of Large to the sample of 316 pulsars detected in four major
400-MHz surveys. Two key results from their paper are shown in
Fig.~\ref{fig:lmt}.
As a result of their numerical analysis, LMT were able to place better
constraints on the radial distribution than TM. However, due to the
extreme selection effects on 400-MHz surveys towards the inner regions
of the Galaxy, the corrected radial distribution becomes very
uncertain below 4 kpc (see Fig.~\ref{fig:lmt}).

To derive the birth rate, ${\cal R}$, rather than applying a pulsar
current anlalysis, LMT followed GO's approach to model the observed
distributions and concluded that the best-fit timescale for
exponential field decay to be 9.1 Myr. Since their analysis assumed
GO's $L \propto B^2$ luminosity law, magnetic field decay provided a
mechanism for luminosity decay. By considering their corrected
luminosity function as a steady-state population where pulsars flowed
into successively fainter luminosity bins, LMT were able to derive
${\cal R}$ as a function of $L$. The result of this analysis is shown
in Fig.~\ref{fig:lmt}, from which LMT concluded ${\cal R} = 1/(30-250
\, {\rm yr})$.

While an elegant approach to the problem, LMTs method relies on the
field decay hypothesis being correct. As mentioned previously, the
issue of magnetic field decay remains unresolved. Since magnetic field
evolution underpins much of pulsar statistics, perhaps the single most
important breakthrough in this area would be a comprehensive
reassesment of this issue.

\section{Recent progress in the Galactic distribution of pulsars}

Although much effort has gone into improving the Monte Carlo
simulations of pulsar population modeling since LMT, relatively
little progress has been made in improving our knowledge of the
Galactic distribution.  Of particular interest is the underlying
density $\rho_R$ of pulsars as a function of Galactocentric radius,
$R$. LMT and others were only able to place poor constraints on this
function (Fig.~\ref{fig:lmt}) and most subsequent work has assumed a
Gaussian distribution for $\rho_R$ (e.g.~Narayan 1987). In fact,
as pointed out by Bailes \& Kniffen (1991), there is no reason to
prefer this function over one where $\rho_R \rightarrow 0$ as
$R \rightarrow 0$.

Improving our understanding of $\rho_R$ at small $R$ requires better
statistics of the inner-Galaxy pulsars.  Due to the propogation and
sky-background selection effects mentioned earlier, low-frequency (0.4
GHz) surveys are very poor probes of this population. One of the main
motivations for high-frequency (1.4 GHz) surveys of the Galactic plane
is that they are less prone to these effects.  Johnston (1994)
analysed two such surveys carried out in the late 1980s (Clifton et
al.~1992; Johnston et al.~1992) and found that models with a deficit
of pulsars in the inner Galaxy were strongly preferred over a simple
Gaussian profile for $\rho_R$.

The sample of $\sim 150$ pulsars detected by these two early surveys
has now been completely surpassed by the phenomenal success of the PM
survey: over 700 new pulsars have been found in the Galactic plane
search so far. Together with re-detections of known pulsars, the PM
sample amounts to 914 pulsars.  I have applied Large's numerical
approach to this new sample
\begin{figure}[hbt]
\psfig{file=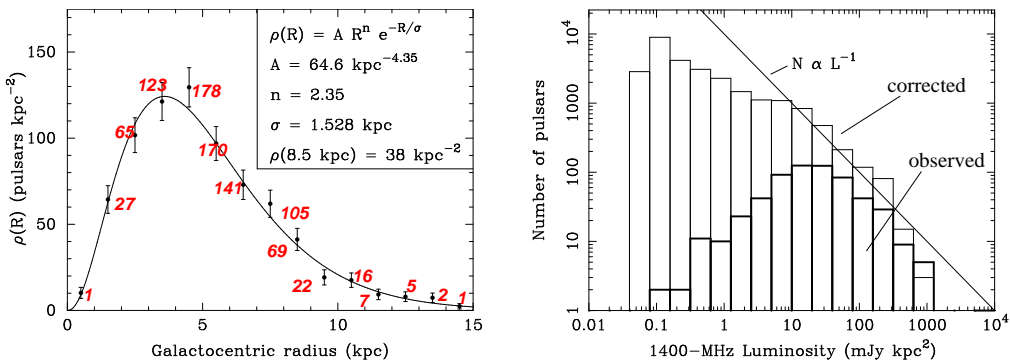,width=\textwidth}
\caption{\label{fig:new} Left: The corrected radial
distribution from the PM survey. 
The inset shows the best-fit analytical function to the data.
The number of observed pulsars 
used to constrain the density in each bin are given. 
Right: The observed and corrected luminosity distributions.}
\end{figure}
to derive new distribution functions in $R$, $L$, $z$ and $P$.
Preliminary results are shown in Fig.~\ref{fig:new}.  The corrected
radial density function clearly supports Johnston's conclusion for a
deficit of pulsars in the inner Galaxy. More work is required in
quantifying the significance of this result, particularly in the
$R=0.5$ kpc bin where only one pulsar is presently known!  While a
population of inner-Galaxy pulsars could be masked from the PM survey
by severe scattering not currently taken into account, it is
interesting that the now completed Effelsberg 5-GHz Galactic centre
survey (Klein et al.~these proceedings) has not found a single pulsar.

Although the PM survey has been the most prolific probe of
the Galactic population to date, we are still only scratching
the surface. One of many uncertain areas about the pulsar population
is the shape of the luminosity function. From Fig.~\ref{fig:new}
we see a clear departure from the $d \log N/d \log L = -1$ 
relationship at low $L$. Integrating Fig.~\ref{fig:new} results in a 
Galactic population of $25,000 \pm 2000$ potentially observable 
pulsars with $L_{1400} > 0.3$ mJy kpc$^2$. Below this limit,
the population is essentially unknown. Future surveys with the
SKA, which should easily be able to detect $\sim 15,000$ pulsars
with $L_{1400} > 0.3$ mJy kpc$^2$ should also detect many
fainter objects and truly constrain the pulsar luminosity function.

\acknowledgments
I thank the Particle Physics and Astronomy Research Council
and the Royal Society for supporting my attendance at this meeting.

\end{document}